\documentclass[reprint,a4paper,aps,prl,floatfix]{revtex4-1}
\usepackage{times,amsmath,amsfonts,amssymb,epstopdf,xspace}
\usepackage{graphicx}
\usepackage{hyperref}
\usepackage[usenames]{color}

\newcommand{\eagle}{\textsc{eagle}\xspace}

\newcommand{\apostle}{\textsc{apostle}\xspace}

\newcommand{\mnras}{MNRAS}

\newcommand{\aap}{A\&A}

\newcommand{\apjl}{ApJL}

\def\ltsima{$\; \buildrel < \over \sim \;$}
\def\simlt{\lower.5ex\hbox{\ltsima}}
\def\gtsima{$\; \buildrel > \over \sim \;$}
\def\simgt{\lower.5ex\hbox{\gtsima}}

\begin{document}

\title{The Mass-Discrepancy Acceleration Relation: A Natural Outcome of Galaxy Formation in Cold Dark Matter Halos}

\author{Aaron D. \surname{Ludlow}}
\email[Electronic address: ]{aaron.ludlow@durham.ac.uk}
\author{Alejandro \surname{Ben\'{i}tez-Llambay}}
\author{Matthieu \surname{Schaller}}
\author{Tom \surname{Theuns}}
\author{Carlos S. \surname{Frenk}}
\author{Richard \surname{Bower}}
\affiliation{Institute for Computational Cosmology, Department of Physics,
  Durham University, Durham DH1 3LE, U.K.}

\author{Joop \surname{Schaye}}
\affiliation{Leiden Observatory, Leiden University, PO Box 9513, 2300 RA Leiden, the Netherlands}

\author{Robert A. \surname{Crain}}
\affiliation{Astrophysics Research Institute, Liverpool John Moores University, 146 Brownlow Hill, Liverpool, L3 5RF}

\author{Julio F. \surname{Navarro}}
\altaffiliation{Senior CIfAR fellow}
\author{Azadeh \surname{Fattahi}}
\author{Kyle A. \surname{Oman}}

\affiliation{Department of Physics and Astronomy, University of Victoria,
  PO Box 1700 STN CSC, Victoria, BC, V8W 2Y2, Canada}

\date{\today}

\begin{abstract}
\noindent We analyze the total and baryonic acceleration profiles of a set 
of well-resolved galaxies identified in the \eagle suite of hydrodynamic simulations.
Our runs start from the same initial conditions but adopt different prescriptions
for unresolved stellar and
AGN feedback, resulting in diverse populations of galaxies by the present 
day. Some of them reproduce observed galaxy scaling relations, while others do not.
However, regardless of the feedback implementation, all of our galaxies 
follow closely a simple relationship between the total and baryonic acceleration profiles, 
consistent with recent observations of rotationally supported galaxies. The relation
has small scatter: different feedback implementations -- which produce different 
galaxy populations -- mainly shift galaxies along the relation, rather than 
perpendicular to it. Furthermore, galaxies
exhibit a characteristic acceleration, $g_{\dagger}$, above which baryons dominate the mass
budget, as observed. These observations, consistent with simple modified Newtonian 
dynamics, can be accommodated within the standard cold dark matter paradigm.

\end{abstract} 

\maketitle

In the cold dark matter (CDM) cosmological model 
structures form hierarchically through merging and 
smooth accretion \cite[e.g.,][]{Wang2011}.
The resulting ``dark matter (DM) halos'' trap gas which cools and forms stars,
providing visible tracers of the underlying DM density field \cite{White1978,White1991}. 
Understanding the connection between galaxies and their halos is therefore
of fundamental importance to galaxy formation models. 

Galaxy formation occurs over a broad range of 
scales, which hampers theoretical progress. Even the most
sophisticated numerical simulations available are unable to resolve
all relevant scales simultaneously, and must resort to ``sub-grid''
models that account for unresolved physical process, such as 
feedback from stars and black holes
\cite[e.g.,][]{Guedes2011,Vogelsberger2014,Schaye2015,Grand2017}.
Sub-grid models are ubiquitous in areas of science that probe 
multi-scale phenomena. They are essential
ingredients in, for example, climate or atmospheric models, 
and simulations of turbulent flows.

Traditionally the link between galaxies and halos has been expressed in terms 
of scaling relations between their structural properties; the Tully-Fisher 
\cite[][TF]{TullyFisher1977} and Faber-Jackson \cite[]{FaberJackson1976} relations, 
in particular, relate the luminosity (or stellar mass) of a galaxy to its dynamics which, 
in CDM, is largely governed by its DM halo. 
Galaxy formation models based on CDM do not reproduce these relations 
unless sub-grid models for unresolved feedback are calibrated to form
realistic galaxies when 
judged according to other diagnostics \cite[e.g.,][]{Lacey2016,Ferrero2017}.
It comes as a surprise, then, that observations reveal an even closer coupling
between the luminous mass of galaxies and their total {\em dynamical} mass.
Perhaps most unexpected is the ``mass discrepancy-acceleration relation'' 
(MDAR) \cite{McGaugh2004,Janz2016,Lelli2017}, a tight empirical relation
between the radial dependence of the enclosed baryonic-to-dynamical mass ratio
and the baryonic acceleration.
It has small intrinsic scatter and holds for galaxies of widely varying luminosity 
and gas fraction. The MDAR may be expressed empirically as \cite{McGaugh2016}
\begin{equation}
  \frac{g_{\rm tot}(r)}{g_{\rm bar}(r)}=\frac{{\rm M_{tot}}(r)}{{\rm M_{bar}}(r)}=\frac{1}{1-e^{-\sqrt{g_{\rm bar}/g_{\dagger}}}},
  \label{eq1}
\end{equation}
where $g_i(r)$ and ${\rm M}_i(r)$ are, respectively, the acceleration and
enclosed mass profiles. 

\begin{figure*}
  \includegraphics[width=0.95\textwidth]{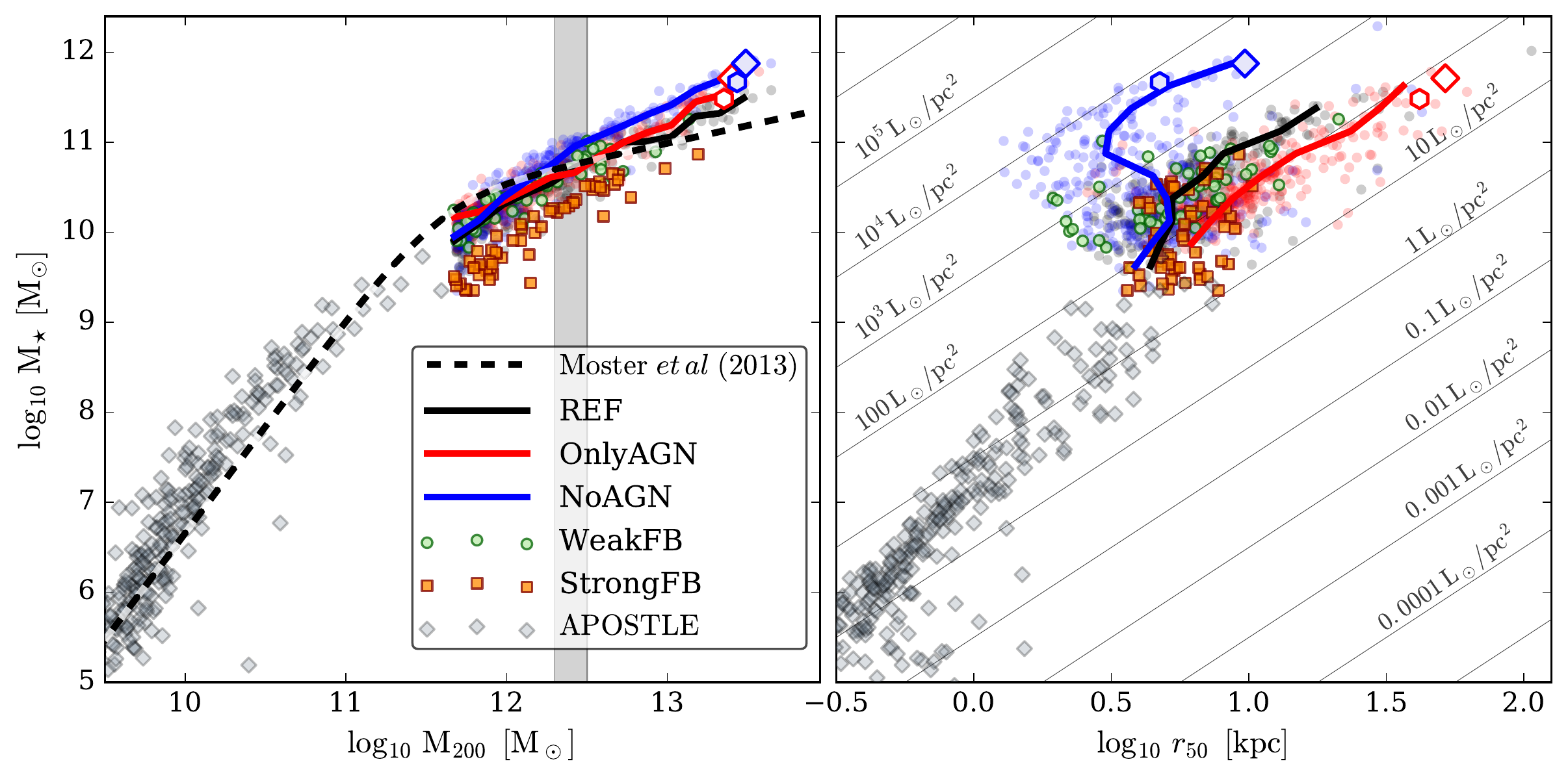}
  \caption{Stellar mass versus halo virial mass (left) and stellar
    half-mass radius (right).
    Solid black lines show the median trends for the ``REF'' model; blue and red lines
    show, respectively, the variations if feedback is entirely limited
    to AGN (OnlyAGN) or to stars (NoAGN). Semi-transparent dots of the same color show individual
    halos. Individual halos are also shown for runs with strong (StrongFB, orange squares) and
    weak (WeakFB, green circles) stellar feedback, and for \apostle galaxies (diamonds).
    The dashed line in the left panel plots the ${\rm M}_\star - {\rm M_{halo}}$ relation
    derived from abundance matching \cite[see][for details]{Moster2013}.
    Lines of constant effective surface brightness are plotted in the right-hand panel. The 
    outsized blue and red symbols identify the two halos shown in Figure~\ref{fig2}.}
  \label{fig1}
\end{figure*}

It has been claimed \cite[see][and discussion therein]{Milgrom2016}
that the small scatter in the MDAR is inconsistent with
hierarchical galaxy formation models, in which galaxies
exhibit a broad range of properties even for halos of fixed mass. Furthermore, the MDAR 
implies a {\em characteristic} acceleration ($g_\dagger\approx 10^{-10}\,{\rm m\,s}^{-2}$) above which 
each galaxy's dynamics can be determined by the observed light alone. 

Why would baryons and dark matter ``conspire'' to produce a characteristic 
physical scale? One possibility is that galaxies adhere to {\em modified Newtonian
dynamics} (although see \cite{Kaplinghat2002,Navarro2016}
for an explanation within the CDM framework).  
However, theoretical studies suggest that the
MDAR arises naturally in CDM models of galaxy formation, provided they also
match observed galaxy scaling relations 
\cite{vdbDalcanton2000,DiCintio2016,SantosSantos2016,KellerWadsley2016,Desmond2016}.
In this letter we address these issues using a suite of simulations drawn from the 
\eagle Project \cite{Schaye2015}. Our simulations vary the subgrid
feedback in a way that modifies the 
end product of galaxy formation, enabling us to robustly assess the
MDAR for a range of galaxy formation ``models''. 

{\em The \eagle Simulations.}--- Our analysis focuses on halos and their central galaxies
identified in a subset of the ``intermediate resolution'' \eagle simulations 
\cite{Schaye2015,Crain2015}. These include periodic volumes of side-length $L_{\rm cube}=25$ and $50$
comoving Mpc sampled with, respectively, ${\rm N}=376^3$ and $752^3$ particles
of gas and DM. The respective particle masses are
$m_{\rm g}=1.81\times 10^6{\rm M}_\odot$ and $m_{\rm dm}=9.70\times 10^6{\rm M}_\odot$;
the (Plummer-equivalent) softening length is $\epsilon=0.7$ physical kpc below $z=2.8$, and
2.66 comoving kpc at higher redshift. 
Each volume was also carried out using only DM, with
$\Omega_{\rm M}'=\Omega_{\rm M}+\Omega_{\rm bar}$
and $\Omega_{\rm bar}'=0$. In all runs, DM particles were assigned unique
integer IDs; we use the same IDs for particles in runs that start from
the same ICs. DM halos can then be matched across different simulations 
by identifying halos with common DM particles.
Cosmological parameters are those inferred by the 
Planck Collaboration \cite{Planck2014}.

\begin{figure*}
  \includegraphics[width=0.95\textwidth]{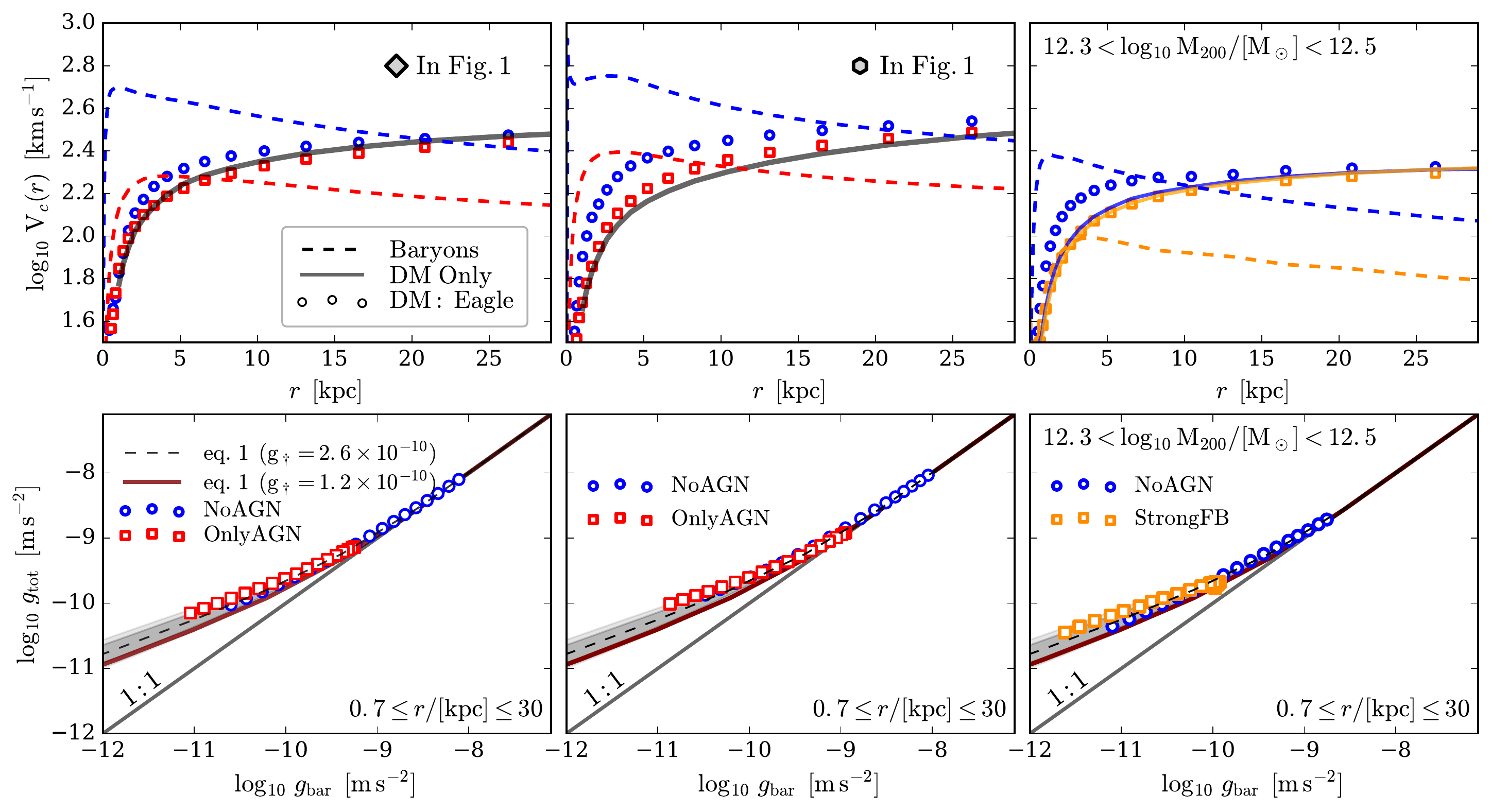}
  \caption{Circular velocity (upper panels) and acceleration profiles (lower panels)
    for galaxies highlighted in Figure~\ref{fig1}. Left and middle panels 
    correspond to individual halos cross-matched between the NoAGN (blue colors)
    and OnlyAGN (red) models; right-most panels compare the {\em median} profiles for 
    halos in NoAGN (blue) and StrongFB (yellow) that fall in the narrow mass range 
    $12.3\leq \log_{10}\, {\rm M_{200}/[{\rm M}_\odot]}\leq 12.5$ (vertical 
    shaded band in Figure~\ref{fig1}).
    The baryonic circular velocity profiles (upper panels)
    are shown using dashed lines; symbols indicate that of DM. 
    (For comparison, solid black lines show the ${\rm V^{DM}_c(r)}$ profiles for 
    the same halo identified in the
    corresponding DM-only simulation.) Lower panels show the acceleration diagrams.
    The linear scaling is shown as a solid black line and 
    eq.~\ref{eq1} (using $g_\dagger=2.6\times 10^{-10}\, {\rm m \,s}^{-2}$) as a dashed line;
    shaded regions indicate the scatter brought about by increasing or decreasing the enclosed
    baryon mass by factors of $3$ (light) and $2$ (dark). For comparison, we also show eq.~\ref{eq1}
    with $g_\dagger=1.2\times 10^{-10}\, {\rm m \,s}^{-2}$ (brown line), consistent with the 
    observational result of \cite{McGaugh2016}.} 
  \label{fig2}
\end{figure*}

The simulations were performed with a version of the N-body hydrodynamics code 
{\sc gadget3} \cite{Springel2005b} incorporating
a modified hydrodynamic scheme, time-stepping criteria and subgrid physics modules
\cite[see][for details]{Schaye2015}. Runs of a given boxsize
start from the same initial conditions but adopt 
different values of the subgrid parameters.
As a result, some accurately
reproduce a diverse set of observations of the galaxy population
(such as the stellar mass function, galaxy shapes and their relationship 
to stellar mass), whereas others do not.

As discussed by \cite{Schaye2015}, 
calibration of the subgrid parameters must be carried out so that
simulations reproduce a diagnostic set of observational
data. For \eagle, this was achieved by calibrating the
feedback models (including contributions from both active galactic
nuclei, AGN, and stars) so that the observed galaxy stellar mass
function and the mass-size relation were recovered. One such model is
the ``reference'' model (hereafter REF; \cite{Schaye2015}).
Variations of REF systematically changing the subgrid
parameters were also carried out \cite{Crain2015}. These include runs
with weak (WeakFB) or strong (StrongFB) stellar feedback, one with
no AGN feedback (NoAGN), and another with {\em only} AGN
feedback but none from stars (OnlyAGN). The resulting galaxy properties 
depend sensitively on these feedback choices.

{\em Analysis: Halo Finding and Selection.}--- We use {\sc subfind} \cite{Springel2001b,Dolag2009} to 
identify DM halos and their {\em central} galaxies (see \cite{Schaye2015} for details).
The position of the halo particle with the minimum potential energy 
defines the halo and galaxy center. 
The halo's virial mass, ${\rm M_{200}}$, is defined as that enclosed by a
sphere of mean density $200\times \rho_{\rm crit}$ surrounding each halo center,
where $\rho_{\rm crit}=3{\rm H_0}^2/8\pi{\rm G}$ is the critical density. 
This implicitly defines the virial radius 
through ${\rm M}_{200}=(800/3)\pi \, r_{200}^3\, \rho_{\rm crit}$.
We focus our analysis on central
galaxies whose DM halos are resolved with at least ${\rm N(<r_{200})}\geq 5\times 10^4$ 
particles. We impose no isolation or relaxation criteria. 

We also include {\em isolated} galaxies (that lie beyond
$2\times r_{200}$ from any halo with ${\rm M_{200}}>5\times 10^{11}{\rm M}_\odot$,
but within 3 Mpc from their barycenter) identified in the level-1  
\apostle simulations (see \cite{Sawala2016,Fattahi2016} for
details of the \apostle Project), which used the \eagle subgrid model 
with REF parameters.
In total, our galaxies span the (stellar) mass range
$10^5\simlt {\rm M/M}_\odot\simlt 10^{12}$.

{\em Radial Mass profiles of Baryons and Dark Matter.}--- The acceleration profile due to component $i$ is computed as
\begin{equation}
  g_i(r)=\frac{{\rm G}\,{\rm M}_i(r)}{r^2}\equiv \frac{{\rm V}_c^i(r)^2}{r},
  \label{eq:gacc}
\end{equation}
where ${\rm V}_c^i(r)$ and ${\rm M}_i(r)$ are the corresponding circular
velocity and enclosed mass profiles, and ${\rm G}$ is Newton's constant.
We compute ${\rm M}_i(r)$ using logarithmically-spaced radial bins with
fixed separation, $\Delta\log_{10} r=0.1$, spanning $r_{\rm min}=\epsilon$ 
(the minimum resolved spatial scale) to
$r_{\rm max}= 0.15\times r_{200}$ (this aperture encloses, on average, 
$\simgt$95\% of a galaxy's stellar mass; we have verified 
that our results are robust to reasonable changes in $r_{\rm max}$).

For each galaxy we also record a few diagnostic quantities. Its stellar mass,
${\rm M}_\star$, is defined as the total mass of stars {\em gravitationally bound} to the central
galaxy; the stellar half-mass radius, $r_{50}$, is defined by 
${\rm M}(r_{50})/{\rm M}_\star=1/2$.

\begin{figure*}
  \includegraphics[width=1.\textwidth]{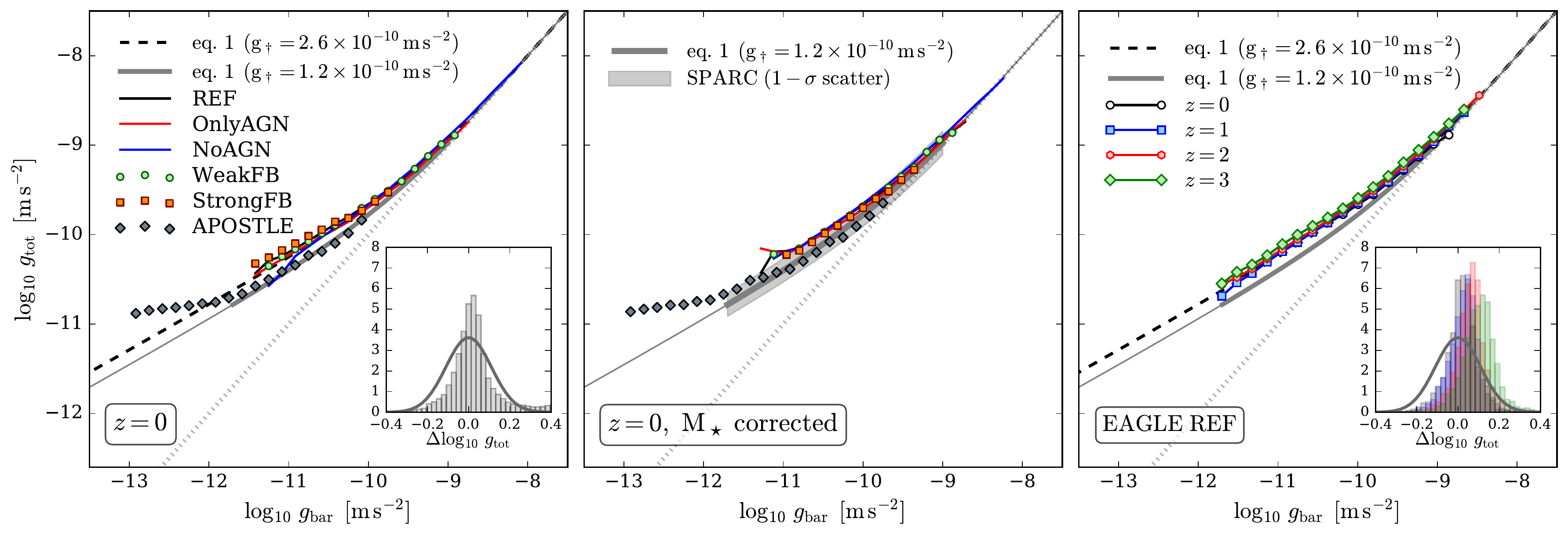}
  \caption{Total acceleration profiles for all halos as a function
    of their baryonic acceleration. The left panel shows results 
    for all halos in all simulations at $z=0$. Lines, points and 
    colors have the same meaning as in Figure~\ref{fig1}. 
    The right-hand panel shows (for REF) the redshift evolution for 
    progenitor galaxies. The dashed lines in 
    the left and right-hand panels show 
    eq.~\ref{eq1} with $g_\dagger=2.6\times 10^{-10}\, {\rm m\, s}^{-2}$.
    Inset panels show the relative scatter around this curve after 
    combining all simulations (left) and for individual redshifts (right); 
    the solid lines represent the observational scatter in 
    \cite{McGaugh2016}. 
    The middle panel plots the $g_{\rm obs}-g_{\rm bar}$ relation after
    rescaling galaxy stellar masses so that they fall on the 
    abundance matching relation shown in Figure~\ref{fig1} (left).
    The thick grey line and shaded band indicates 
    the mean trend and scatter obtained by \cite{McGaugh2016} from 
    observations of rotationally supported galaxies.}
  \label{fig3}
\end{figure*}

{\em Results.}--- The leftmost panel of Figure~\ref{fig1} plots galaxy stellar mass versus halo virial mass.
Solid lines show the median trends for the 50 Mpc cubes (REF, NoAGN and OnlyAGN).
Individual galaxies are shown as faint circles of corresponding color. 
Additional runs with strong and weak stellar feedback are also shown, along with
\apostle galaxies (in these cases only 
individual halos are plotted). The dashed line shows the relation inferred from 
abundance matching on data by \cite{Moster2013}. The right-hand panel shows, using
the same color scheme, the stellar mass versus half (stellar) mass radius. The diagonals
indicate lines of constant surface brightness.

Different subgrid models produce different galaxy populations.
For a given halo mass 
the median galaxy {\em stellar mass} spans a factor of $\approx 4$ between 
the extremes (compare NoAGN and StrongFB in the left-most panel).
Galaxy sizes also differ, particularly for the runs without
(NoAGN) and with {\em only} (OnlyAGN) AGN feedback. For 
${\rm M_{\star}}\simgt 10^{11} {\rm M}_\odot$, for example, galaxy half-mass radii 
are roughly an order of magnitude smaller when AGN feedback is ignored. 

Figure~\ref{fig2} (upper panels) provides a few examples of the circular velocity profiles
of baryons (dashed curves) and dark matter (open symbols) for several \eagle galaxies.
The left and middle panels
show two massive galaxies that were cross-matched in the NoAGN and OnlyAGN runs
(highlighted as outsized points in Figure~\ref{fig1}). 
Because they inhabit the same
halo their merger histories are similar, but their stellar masses,
and sizes differ noticeably as a result of differing feedback processes. 
Each galaxy's DM distribution reflects its 
response to galaxy formation: the more massive the central galaxy, the more 
concentrated its DM halo. The effect is, however, weak. The dark grey line in each
panel shows, for comparison, the circular velocity curve of the same halo
in the corresponding DM-only simulation. 

The resulting rotation curves show a clear transition from baryon to dark 
matter dominated regimes, suggesting that careful calibration of subgrid models is
needed to produce galaxies with realistic mass profiles. Despite these structural 
differences, all four galaxies nevertheless follow
closely the same relation between the total acceleration
and the acceleration due to baryons (lower panels). Galaxies 
in the NoAGN run, which are more massive
and more compact than those in OnlyAGN, populate the high acceleration regime of the 
relation, indicating that they are baryon dominated over a larger radial 
extent. When included, AGN feedback periodically quenches star formation resulting 
in less compact and lower mass central galaxies that are DM dominated over a large radial 
range. 

The right-hand panels of Figure~\ref{fig2} show another example. Here
we select all halos from NoAGN and StrongFB whose masses lie 
in the range $12.3\leq \log_{10} {\rm M_{200}/{\rm M}_\odot\leq 12.5}$ 
(vertical shaded band in the left panel of Figure~\ref{fig1}) and plot their 
{\em median} circular velocity and acceleration profiles. These galaxies
have stellar masses that differ, on average, by a factor of $\approx 4$ depending 
on the feedback implementation, but inhabit halos of comparable DM mass. As before, 
solid curves show the median dark matter mass profile for the same halos 
identified in the corresponding DM-only
simulation; open symbols show ${\rm V_c^{DM}}(r)$ measured directly in the \eagle runs. 
The suppression of star formation by strong feedback results in considerably 
less massive galaxies that are dark matter dominated at most 
resolved radii. Nevertheless, both sets of galaxies follow the 
acceleration relation given by eq.~\ref{eq1}. 

In all cases, {\em different feedback models produce galaxies that move
 along the MDAR rather than perpendicular to it}, resulting in small scatter.
It is easy to see why. Consider an arbitrary galactic radius at which the total
and baryonic accelerations are related by eq.~\ref{eq1}. Changing the enclosed
baryon mass within this radius by a factor $f$ shifts points horizontally 
to $g^{'}_{\rm bar}=f\, g_{\rm bar}$, but also vertically to 
$g^{'}_{\rm tot}=g_{\rm tot}+(f-1)\, g_{\rm bar}$. As a result, galaxies
of different stellar mass or size that inhabit similar halos tend to move
diagonally in the space of $g_{\rm bar}$ versus $g_{\rm tot}$. The
shaded regions in the lower panels of Figure~\ref{fig2} indicate the 
scatter expected for enclosed baryon masses that differ from 
eq.~\ref{eq1} (with $g_\dagger=2.6\times 10^{-10}{\rm m\, s^{-2}}$) by factors
of $3$ (light shaded region) and $2$ (darker region). 

Figure~\ref{fig3} (left) shows the total versus baryonic 
acceleration for {\em all} ($z=0$) galaxies in {\em all} simulations. 
For each run we show the {\em average} trends
either as solid lines (REF, OnlyAGN and NoAGN) or heavy symbols (WeakFB, 
StrongFB and \apostle). The dashed line describes the 
numerical data remarkably well, even for models whose subgrid physics
were {\em not} tuned to match observational constraints. The inset panel
plots the residual scatter around this line.
Despite the wide range of galaxy properties it is smaller ($\sigma=0.08$ dex;
see also \cite{KellerWadsley2016}) than 
that of the best available observational data ($\sigma=0.11$ dex), indicated 
by the solid line \cite{McGaugh2016}. 

Note too that the acceleration relation persists at high redshift, where galaxies
are more likely to be actively merging.
The right-hand panel of Figure~\ref{fig3} shows
the acceleration relation for $z=0$ galaxy progenitors in our REF model at four higher
redshifts.
Regardless of $z$, the mean relations are very similar. The residuals are also 
small (inset panel), but show evidence of a slight but systematic redshift dependence.

Eq.~\ref{eq1} describes all simulations remarkably well, provided 
$g_\dagger\approx 2.6\times 10^{-10}\, {\rm m\,s}^{-2}$ (dashed line). This is a factor of 
$\approx$2.2 larger than that obtained by \cite{McGaugh2016} from observations 
of rotationally supported galaxies. This discrepancy reflects the fact that, 
regardless of subgrid parameters, \eagle systematically 
{\em underpredicts} the stellar content of halos near the knee of the 
${\rm M}_\star-{\rm M_{halo}}$ relation, 
where halos are most abundant. As a result, baryonic accelerations, at fixed $g_{\rm tot}$, 
are smaller than observed. The middle panel of Figure~\ref{fig3} shows
the average MDAR for all simulations after rescaling all galaxy masses
to match the ${\rm M}_\star-{\rm M_{halo}}$ derived from abundance matching
\cite{Moster2013}. All runs are now consistent with the observed relation
to within the observational scatter (shown as a thick grey line and 
shaded region).

{\em Discussion and Summary.}--- We analyzed a suite of simulations from 
the \eagle Project that adopt widely varying subgrid
parameters. Some simulations yield populations of galaxies that differ systematically
from observed galaxy scaling relations. Nevertheless, all galaxies follow a simple
relationship between their total and baryonic acceleration profiles, regardless of
the feedback implementation. Different
feedback prescriptions, which result in different galaxy populations, 
cause galaxies to move {\em along} the MDAR rather than perpendicular to it,
yielding small scatter. 

We note, however, that the total to baryonic acceleration relation 
depends {\em slightly but systematically} on the subgrid model. 
For example, the StrongFB and NoAGN models are, at low acceleration, 
noticeably different: the former lies slightly above the best-fitting eq.~\ref{eq1}, 
the latter slightly below. The differences however are small
and within the observational scatter.
The radial acceleration relation given by eq.~\ref{eq1} is, therefore, 
very forgiving: only {\em large} departures from any sensible galaxy-halo scaling 
relations lead to noticeable systematics. The ``small'' observed scatter in 
the MDAR is, in fact, quite large, and is unlikely to provide useful 
constraints on sub-grid models for galaxy formation.

\vspace{0.2cm}
We thank Lydia Heck and Peter Draper, whose technical support and 
expertise made this project possible, and our referees for useful 
reports. ADL is supported by a COFUND 
Junior Research Fellowship; RAC is a Royal Society University Research Fellow. 
JS acknowledges support from the Netherlands Organisation for Scientific
Research (NWO), through VICI grant 639.043.409, and the European
Research Council under the European Union's Seventh Framework Programme
(FP7/2007- 2013) / ERC Grant agreement 278594-GasAroundGalaxies.
This work was supported by the Science
and Technology Facilities Council (grant number ST/F001166/1);
European Research Council (grant numbers GA 267291 ``Cosmiway'').
Computing resources were supplied bt the DiRAC Data Centric system 
at Durham University, operated by the Institute for Computational 
Cosmology on behalf of the
STFC DiRAC HPC Facility (\url{www.dirac.ac.uk}). This equipment was
funded by BIS National E-infrastructure capital grant ST/K00042X/1,
STFC capital grant ST/H008519/1, and STFC DiRAC Operations grant
ST/K003267/1 and Durham University. DiRAC is part of the National
E-Infrastructure. We also acknowledge PRACE for granting us access to the
Curie machine based in France at TGCC, CEA, Bruy\`eres-le-Ch\^atel.

\bibliographystyle{apsrev4-1} 
\bibliography{paper}

\label{lastpage}

\end{document}